\documentclass[onecolumn]{aa}
\usepackage{graphicx}
\newcommand{\lcu}{\mbox{$\, \stackrel{ < }{ _{\sim} } \,$}}

\begin{document}
   \title{Calibrated stellar models for metal-poor populations}

   \author{P. Cariulo
          \inst{1}
          \and
          S. Degl'Innocenti
        \inst{1}$^,$ \inst{2}
      \and
      V. Castellani\inst{3}$^,$ \inst{4}
          }

   \offprints{S. Degl'Innocenti}
    \institute{ Dipartimento di Fisica, Universit\`a di Pisa,
         via Buonarroti 2, 56126 Pisa, Italy
      \and
       INFN Sezione di Pisa, via Buonarroti 2, 56126 Pisa, Italy
       \email{scilla@df.unipi.it}
        \and
         Osservatorio Astronomico di Roma, via Frascati 33, 00040
         Monte Porzio Catone, Italy
        \and
         INFN, sezione di Ferrara, via Paradiso 12, 44100 Ferrara,
         Italy}

   \date{Received 16 June 2003 / Accepted 2 March 2004}

   \abstract{

We extend to lower metallicities recent evolutionary computations
devoted to Magellanic Clouds stars, presenting and discussing new
stellar models with Z=0.0002, 0.0004, 0.0006, 0.001 and suitable
assumptions about the original He content. As in the previous
paper, evolutionary results are compared with
observational data to properly calibrate the assumptions
about the efficiency of the surface convection. On this basis, we
follow the evolution of stellar models in the mass range 0.6 to
11M$_{\odot}$ from the Main Sequence (MS) to the C ignition or
the onset of thermal pulses in the advanced Asymptotic Giant
Branch (AGB) phase, presenting cluster isochrones covering the
range of ages from 20 Myr to 20 Gyr. Selected predictions
constraining the cluster ages are discussed, presenting a
calibration of the difference in magnitude between the luminous MS
termination and the He burning giants in terms of the cluster age.
Both evolutionary tracks and isochrones are available
at the URL http://astro.df.unipi.it/SAA/PEL/Z0.html\footnote{Data files 
are also available at the CDS}.

\keywords{stars:evolution, globular clusters:general, open clusters and
associations:general
               }
   }

   \maketitle
%

\section{Introduction}
In modern astrophysics, stellar clusters play the role of cosmic
clocks marking the history of star populations in the Galaxy.
Beyond the Galaxy, improved observational capability has already
opened the study of stellar clusters in nearby galaxies, such as
the Small and Large Magellanic Clouds, giving access to the
history of star formation in the Universe. However, a precise
determination of the ages of stellar clusters is still a problem
addressed by several authors, investigating  both the old Galactic
globulars (see e.g. Chaboyer 2001, Salaris \& Weiss 2002) and the
more recent disk  population (see e.g. Carraro et al. 1999,
Chaboyer et al. 1999).  Thus, there is the need for extensive 
evolutionary calculations covering
suitable grids of chemical compositions and ages.  

In a previous paper
(Castellani et al.  2003, Paper I) we presented evolutionary results
for the Magellanic Cloud (LMC, SMC) 
chemical composition, prepared for the analysis
of new HST observations of the intermediate-age LMC globular NGC1866
(Brocato et al. 2003).  The investigation of the deep color-magnitude (CM)
diagram of stellar populations in the Carina dwarf galaxy (Monelli et
al. 2003) has recently suggested the extension of such evolutionary
analysis to lower metallicities. In this paper we present these new
results, discussing the justification for several choices made when 
constructing these stellar models and comparing the results with
similar computations available in the recent literature.

The reasons for presenting new theoretical computations have been
discussed in Paper I, so they will only be stated briefly here. As a
central point, the present set of models is currently the only one
covering the phases of both H and He burning over a large range of
masses with the inclusion of  atomic diffusion of helium and heavy
elements. This process has been demonstrated to sensitively affect
the evolution of low mass stars, with consequences for the
estimated ages of old clusters (see, e.g., Castellani \&
Degl'Innocenti 1999, VandenBerg et al. 2002). We use throughout
the canonical assumption of inefficient overshooting so the He burning
structures are calculated according to the prescriptions of canonical
semiconvection induced by the penetration of convective elements in
the radiative region (Castellani et al. 1985).  In the case of low
mass stars this gives lifetimes for Horizontal Branch (HB) stars in
better agreement with observational constraints than do overshooting
models (see Paper I).

Since the occurrence and/or the amount of overshooting in more massive
stars is still a debated question (see e.g. Bertelli et al. 2003 and
references therein, Brocato et al. 2003), the availability of a set of
canonical metal-poor models will be useful to further
investigate this point.  Moreover, as discussed in Appendix A, the present
models are the only ones based on the most updated physics
available in the literature. This does not necessarily imply that the
updated physical inputs produce the most reliable models, but only
that our models represent an updated approach to the difficult problem
of stellar modeling (see e.g. Castellani et al. 2001).

\section{Model tests and calibrations}
\begin{table*}
\caption{The three clusters selected for isochrone calibrations. For
each cluster the columns give the source of
photometry, the range of reddening values provided in the recent literature, 
E(B-V), [Fe/H] values as given by Zinn \& West (1984:ZW),
Harris (2003:H) and Carretta \& Gratton (1997:CG), the amount of
$\alpha$ enhancement from Salaris \& Cassisi (1996) and the estimates of the
total metallicity, Z, as derived from [Fe/H] 
values for the three quoted references.}
\begin{center}
\begin{tabular}{l c c c c c c c}
\hline
\hline\\
cluster      & photometry       & E(B-V) &$[Fe/H]_{ZW}$ & $[Fe/H]_H$  &  $[Fe/H]_{CG}$& $[\alpha/Fe]$ &Z\\
\hline
NGC4590 (M68) &Walker (1994)    & 0.04$\div$0.09        & -2.09$\pm$0.11 & -2.06  & -1.99$\pm$0.10 & 0.20& 2.0; 2.1;~~ 2.5$\cdot$10$^{-4}$\\
NGC5272 (M3)  &Rey et al. (2001)& 0.01$\div$0.03         &-1.66$\pm$0.06 & -1.57  & -1.34$\pm$0.06 & 0.26& 5.9; 7.3; 12.3$\cdot$10$^{-4}$ \\
NGC6205 (M13) &Rey et al. (2001)& 0.01$\div$0.03        &-1.65$\pm$0.06 & -1.54  & -1.39$\pm$0.06 & 0.28&6.3; 8.1; 11.4$\cdot$10$^{-4}$\\
\hline \hline
\end{tabular}
\end{center}
\end{table*}
The theoretical background and the input physics adopted in our code
have been discussed in Paper I (see also Ciacio et al. 1997
and Cassisi et al. 1998) and are listed in Appendix A.  We only
point out here that our models include atomic diffusion, including the
effects of gravitational settling, and thermal diffusion with
diffusion coefficients given by Thoul et al. (1994) while radiative
acceleration (see e.g. Richer et al. 1998, Richard et al. 2002) is not
taken into account.  For convective mixing, we adopt 
the Schwarzschild criterion  
to define regions in which convection elements
are accelerated (see the description in Brocato \& Castellani 1993).
It is known that diffusion is a slow process; for
masses of about 3 M$_{\odot}$ or higher (that is for clusters of ages
lower that about 350 Myr) its influence is 
negligible.  Effects of rotation (see e.g. Maeder \& Zahn 1998,
Palacios et al. 2003) are not included in our models.

As in previous papers, to assess the adequacy of the predicted
evolutionary scenario we follow the procedure of submitting stellar
models to a preliminary test to calibrate the mixing length parameter
governing the efficiency of convection at the surface of a stellar
structure and, thus, the predicted effective temperature of stars with
convective envelopes.

Following this procedure, our models have been 
satisfactorily tested relative to Solar Standard Models (SSM; see
e.g. Degl'Innocenti et al. 1997), young metal-rich galactic
clusters with Hipparcos parallaxes (Castellani et al. 2001,
Castellani et al. 2002) and moderately metal-rich galactic
globulars (Paper I).  We have adopted -- in all cases -- a ratio $\alpha$
between the mixing length and the local pressure scale height of
$\alpha$ = l/H$_{\mathrm p}$ $\sim 1.9$.  Here we note that SSM
requires element diffusion, so the calibration of $\alpha$ with solar
models without diffusion does not appear meaningful. The same models
also provided a satisfactory fit 
for young globulars in the LMC (Brocato et al. 2003).

For metal-poor stars, one finds in the
Galactic globulars the natural candidates for a further test. In
Cassisi et al. (1999) we fitted our low mass stellar
models to globular clusters  adopting $\alpha$ $\sim 2.0$ 
for metallicities Z$\sim 10^{-4}$ ,
but $\alpha \sim  2.3$ for Z$\sim  10^{-3}$. As we will discuss
in this section, the increased accuracy of available CM diagrams
allows us to submit the above calibration to even more stringent
tests, investigating  in particular the disturbing evidence for the need
for a larger mixing length only for clusters with a metallicity
of about Z$\sim 10^{-3}$.
\begin{figure}
\centering
\includegraphics[width=8cm]{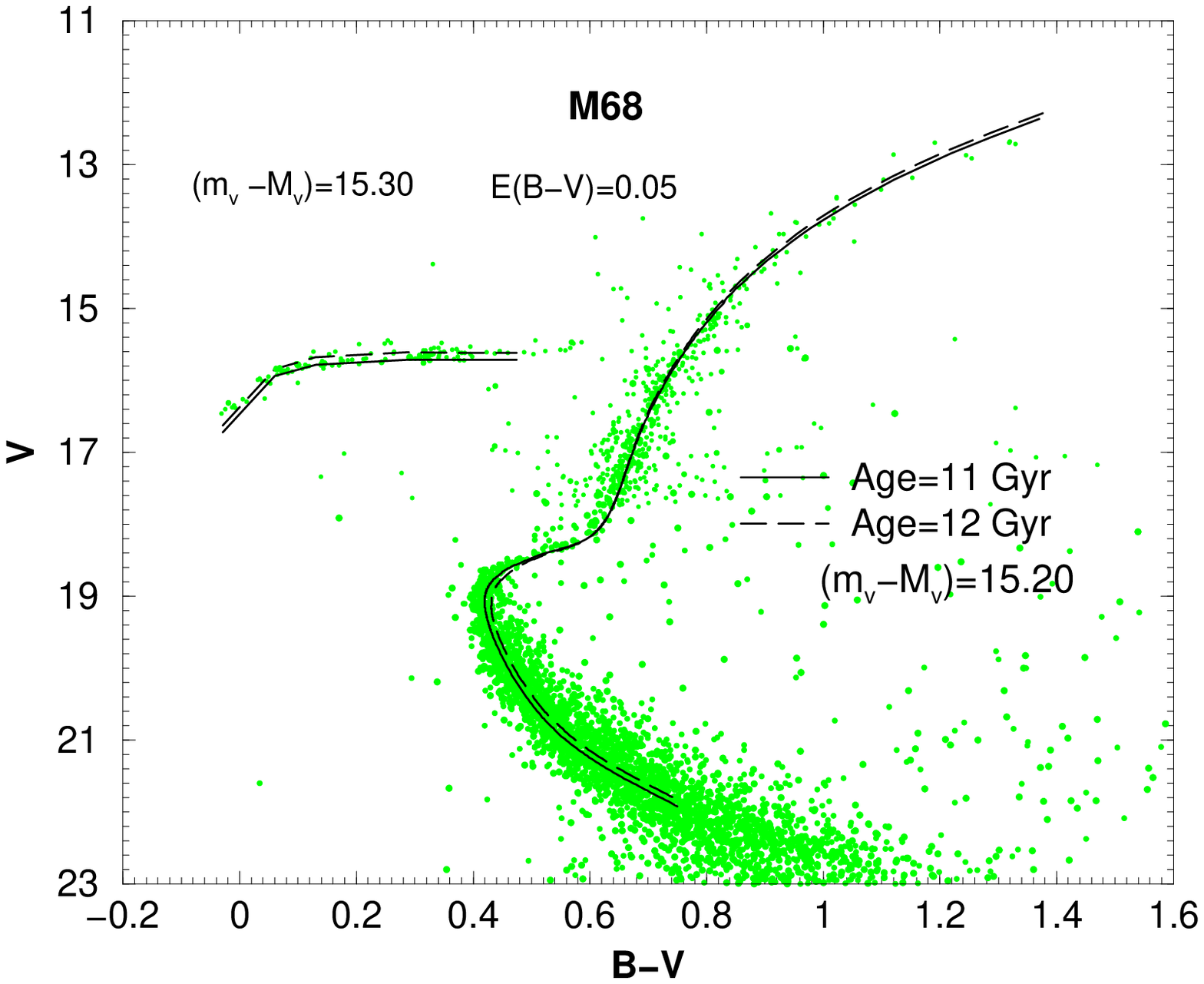}\includegraphics[width=8cm]{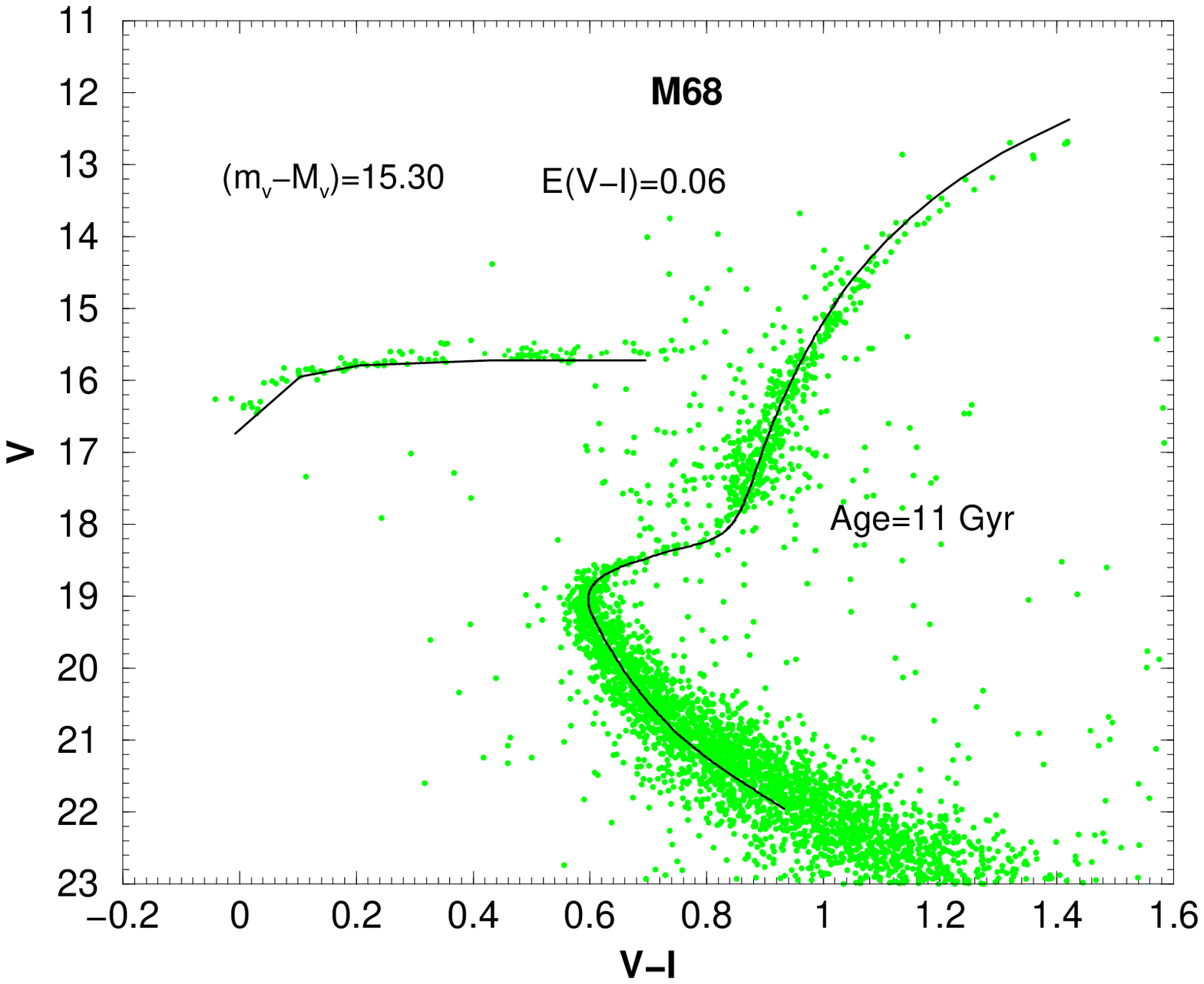}
\caption{The best fit of M68 as obtained with the isochrone for Z=
0.0002, 11 Gyr and mixing length parameter $\alpha$= 2.0 in (V, B-V)
(left panel) and in (V, V-I) diagrams (right panel). In the left panel
the fit with the 12 Gyr isochrone and (m$_{\mathrm v}$-M$_{\mathrm
v}$)=15.20 is also shown (dashed line). Color transformations are from
Castelli (1999, see also Castelli et al. 1997).}
\end{figure}
\begin{figure}
\centering
\includegraphics[width=8cm]{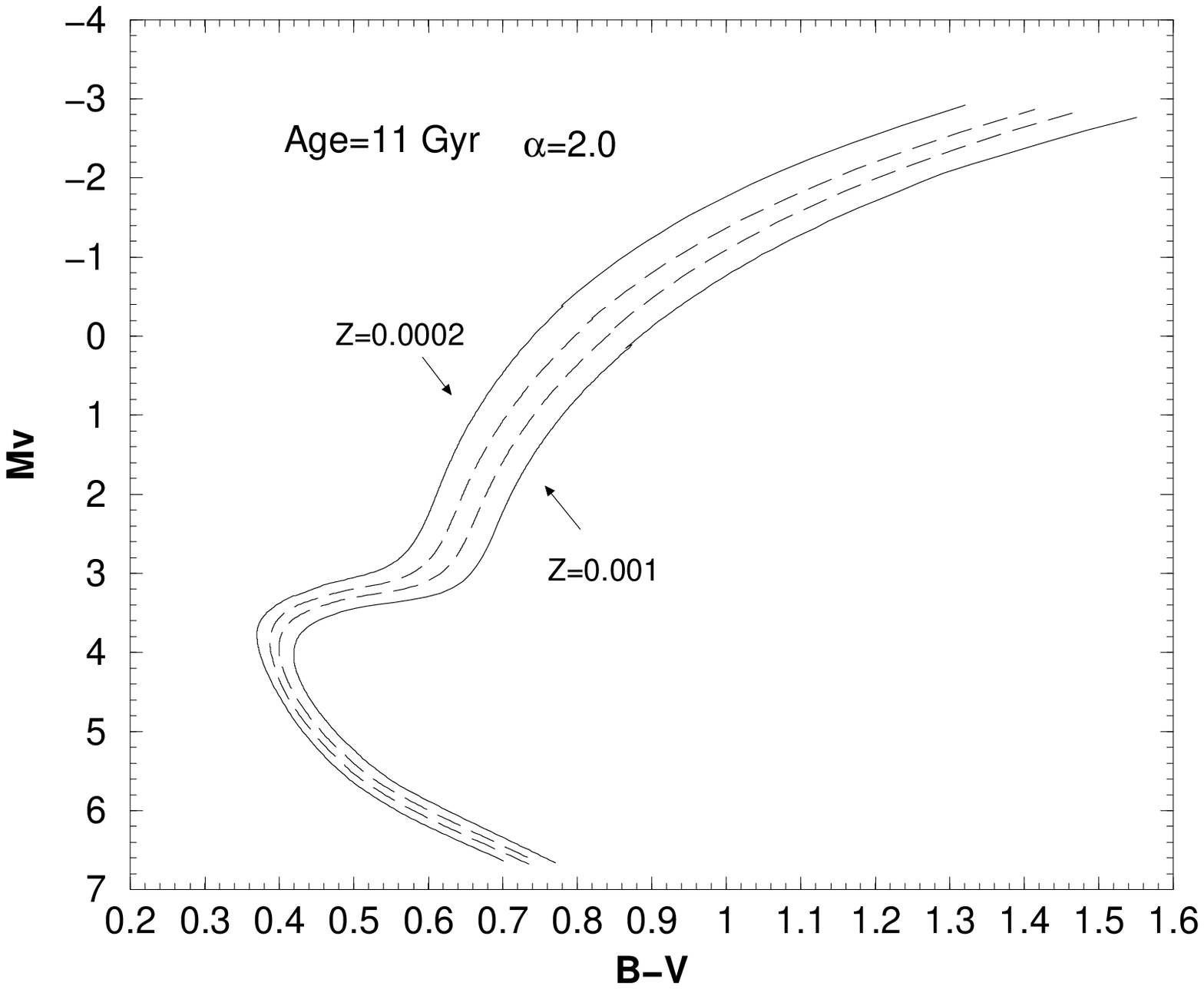}\includegraphics[width=8cm]{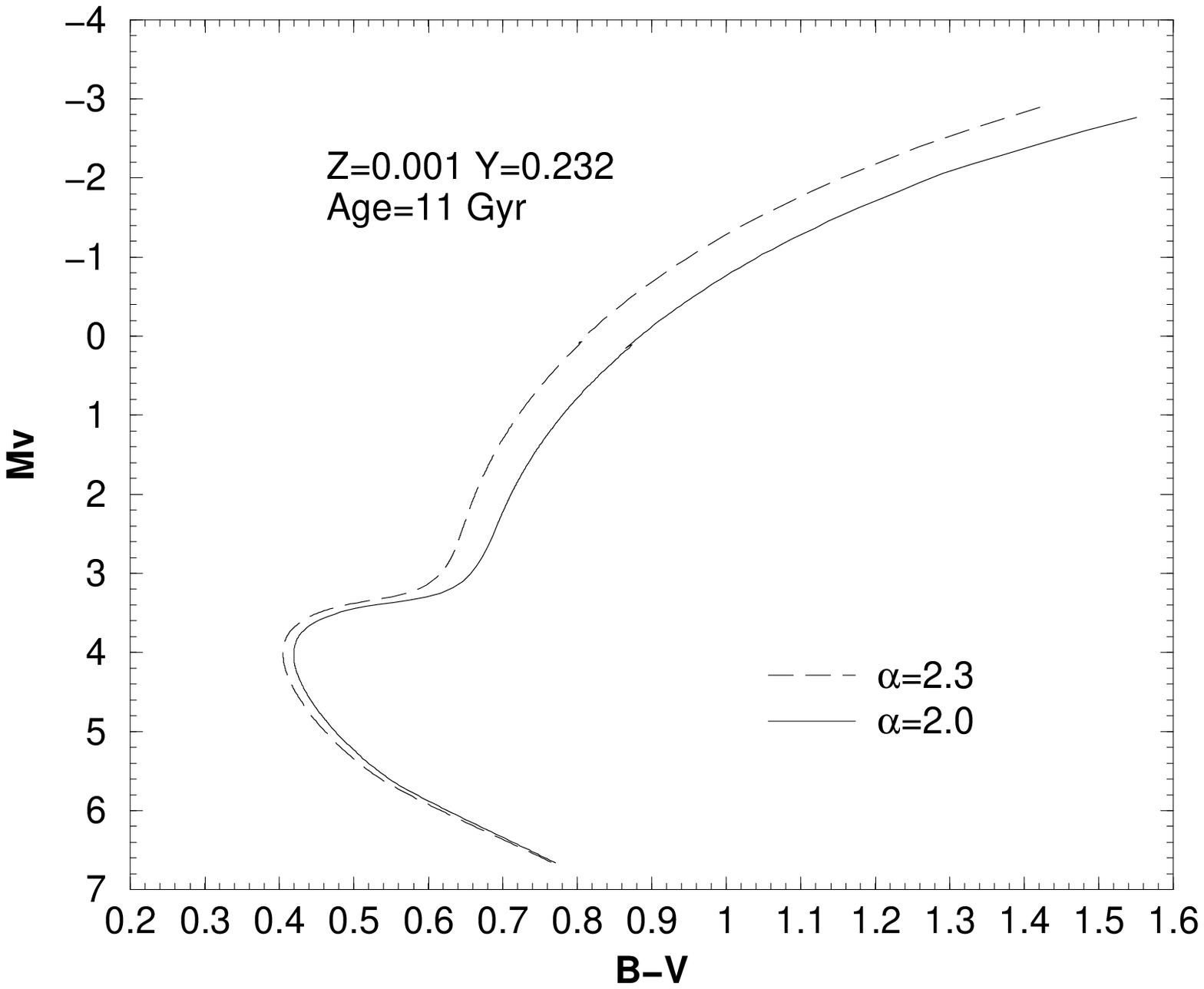}
\includegraphics[width=8cm]{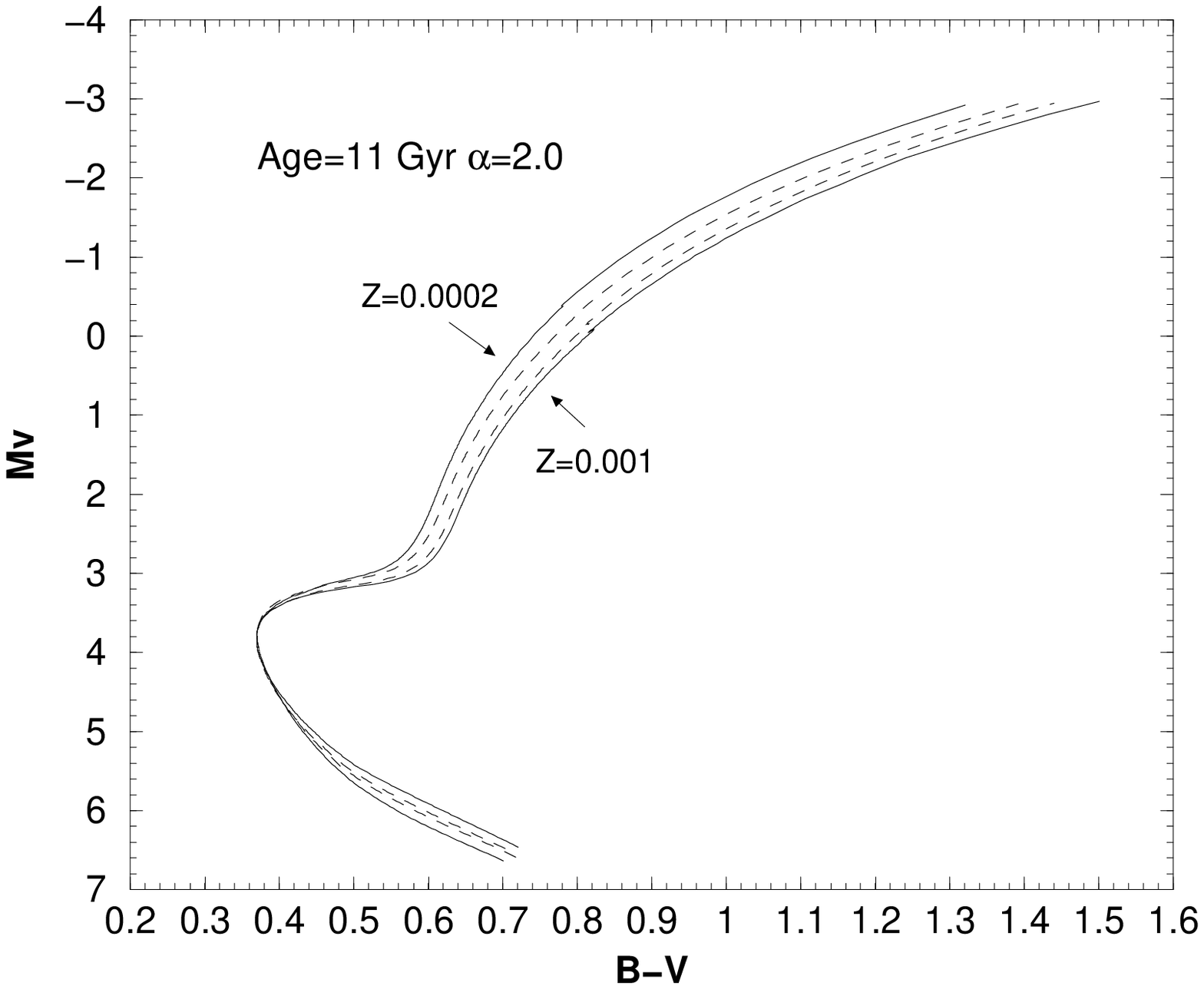}\includegraphics[width=8cm]{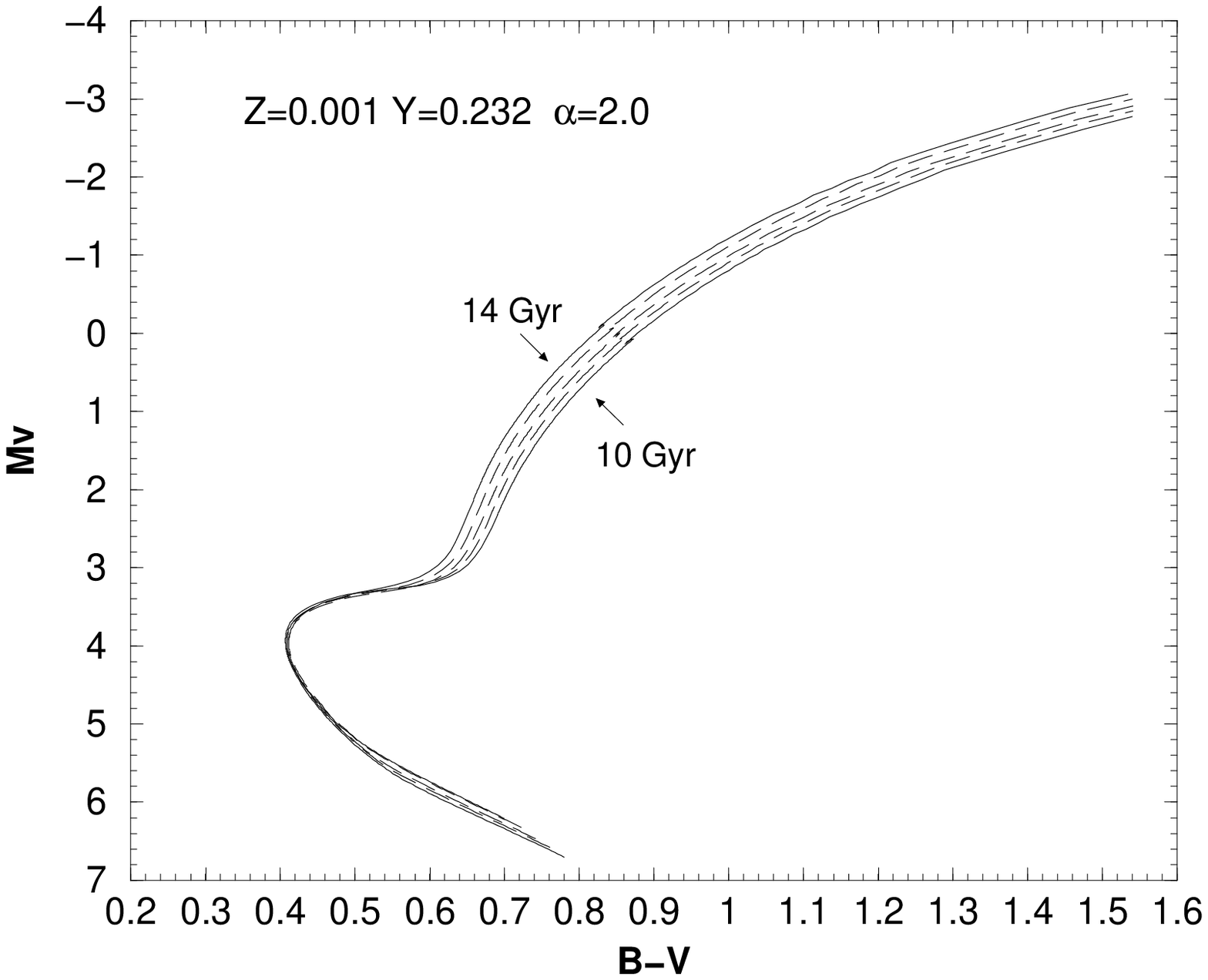}
\caption{Upper left panel: theoretical isochrones for $\alpha$=2.0,
age 11 Gyr and Z= 0.0002, 0.0004, 0.0006, 0.001. Upper right panel:
theoretical isochrones for Z= 0.001, age 11 Gyr, as computed under
the two labeled assumptions about the mixing length. Bottom left
panel: the isochrones as in the upper left panel but shifted to make
TO.s to coincide each other; bottom right panel: isochrones for
Z=0.001 and different ages shifted as in the left panel.  Color
transformations from Castelli (1999, see also Castelli et al. 1997).}
\end{figure}
To this purpose we searched in literature for well observed,
not too heavily reddened globular clusters for which good (V, B-V)
diagrams are available and with [Fe/H] values in the range
explored in this paper.  We consequently selected (as in Cassisi et
al. 1999) M68 as a suitable sample of the most metal-poor
clusters, but now adding the very accurate CM diagrams recently
made available by Rey et al. (2001) for the moderately metal-rich
clusters M3 and M13. Table 1 lists the three selected clusters
together with the source of the photometric data, the range of values 
for the reddening available in the recent literature,
current estimates  for [Fe/H] and for the amount  of $\alpha$-enhancement.

As is well known, the effect of the $\alpha$-enhancement on the
evolutionary tracks and isochrones of Population II stars can be
simulated by using a scaled solar mixture corresponding to the actual
total metallicity. Isochrones calculated with enhanced
$\alpha$ elements are very well mimicked in all the main
characteristics by the standard solar mixture isochrones of
metallicity: Z=Z$_0$(0.638f$_{\alpha}$ + 0.632) where
f$_{\alpha}$=10$^{[\alpha/\mathrm{Fe}]}$ and Z$_0$ is the initial
(nonenhanced) metallicity (see e.g. Salaris et al. 1993, Salaris \&
Weiss 1998). The last column in Table 1 gives these ``equivalent Z"
as evaluated on the basis of [Fe/H] values according to well known
relations and by adopting the recent determination
[Z/X]$_{\odot}$=0.0230 (see e.g.  Bahcall et al. 2001).

Data in Table 1 show that cluster metallicities are in general far
from being firm observational results. For the most metal-poor cluster, M68, 
the uncertainty is still
tolerable and the actual metallicity can be safely put around the
value Z=0.0002. As already found, Fig.1 shows that adopting this
metallicity the best fit of the clusters requires $\alpha$ = 2.0 with
an age of 11 Gyr. As in previous works, the fitting has been performed
making use of the predicted HB luminosity to constrain the cluster
distance modulus and, consequently, the cluster age.  However, one may be
reluctant to rely uncritically on such a theoretical prediction, since
HB models are much more dependent on uncertaintes in the physical
inputs than TO and subgiant branch (SGB) models.  In this context
evidence from the pulsational behavior of RR
Lyrae variables in galactic globulars has already indicated that
perhaps (i.e. if pulsational theory is taken as face value)
our theoretical HB are too bright, even if by only $\approx$0.05 mag
(Marconi et al. 2003). The discrepancy increases for models without
diffusion. However, Fig.1 also shows that the calibration of
$\alpha$ is not dramatically dependent on the assumed cluster distance
modulus.  Decreasing the distance modulus by 0.1 mag to 15.20 one
would derive a cluster age of 12 Gyr with the same
$\alpha$. Here we note that the distance modulus so derived for
the cluster appears in good agreement with
the results by Carretta et al. (2000, but see also Gratton et
al. 1997) who used Hipparcos parallaxes to obtain (m$_{\mathrm
v}$-M$_{\mathrm v}$)=15.25$\pm$0.06.

To further discuss this calibration, we show in Fig.2 
a set of theoretical data displaying the behavior
of theoretical isochrones with varying metallicity, age or $\alpha$.
We find, for example, that increasing $\alpha$ from 2.0 to 2.3 
decreases the extension in temperature of the SGB, the 
best fit requires shorter ages and also a larger distance modulus
to balance the effect of the increased $\alpha$.
Thus, we will keep $\alpha$ = 2.0 as a rather robust calibration of 
the mixing length in our models at Z= 0.0002, bearing in mind that 
this calibration is supported by independent evaluations of the
cluster apparent distance modulus.  As a parallel check,
 we took advantage of the (V, V-I) diagram presented by Walker (1994) for M68 
to further test the theoretical scenario. As shown in Fig. 1, 
one finds that the best fit obtained in the (V, B-V)
plane is preserved when passing to (V, V-I), supporting the
consistency of the adopted model atmospheres by Castelli (1999).
\begin{figure}
\centering
\includegraphics[width=8cm]{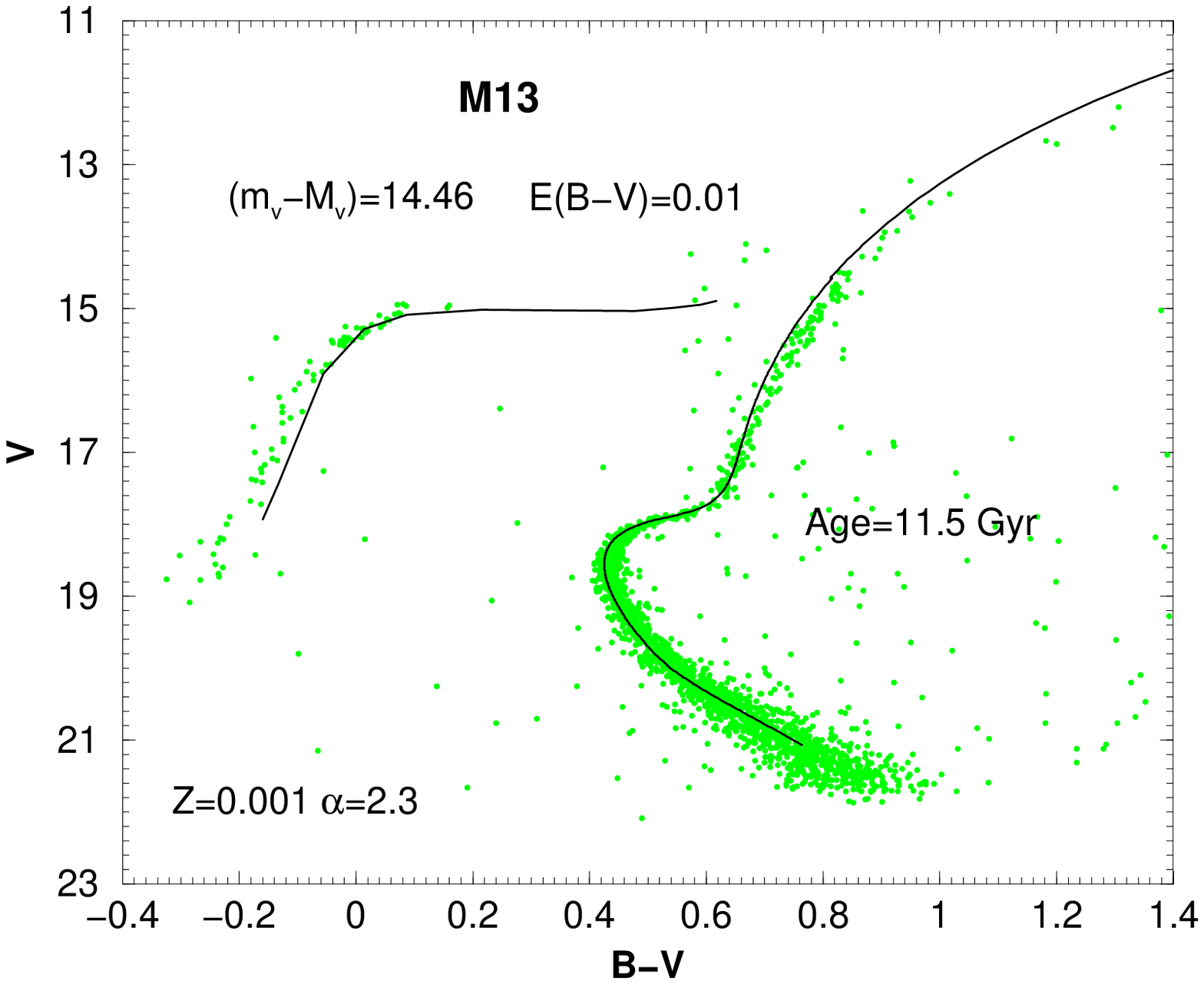}\includegraphics[width=8cm]{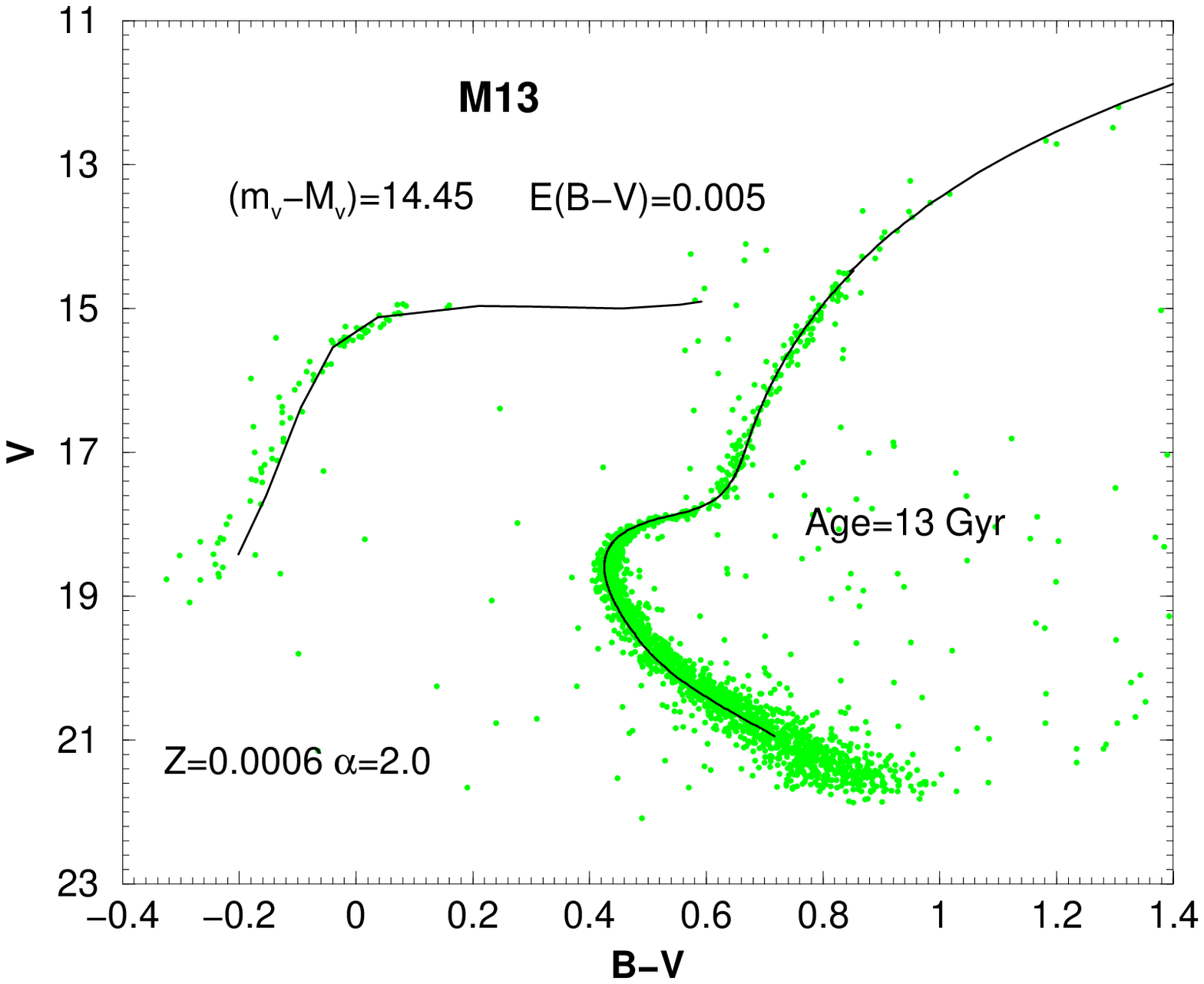}
\caption{Left panel: the best fit of M13 as obtained with the isochrone for 
Z=0.001, 11.5 Gyr and mixing length parameter $\alpha$=2.3.
Right panel: the same but for Z=0.0006, 13 Gyr and $\alpha$= 2.0 (see text).} 
\end{figure}
When passing to larger metallicities, data in Table 1 show that,
unfortunately, the uncertainties in the metallicity of the two
calibrating clusters M3 and M13 are large, while the calibration
depends sensitively on the assumption about this parameter.
The only results on which all authors agree is that the metallicity
of these two clusters should be very similar.
Rey et al. (2001) showed that the subgiant branch in M13 appears
shorter than in M3, in spite of the assumed common metallicity,
reaching the conclusion that M13 is older than M3 by about 2 Gyr.  
A comparison between the Rey et al. data for the two
clusters unambiguously reveals that- if both clusters have a common
metallicity - M3 has a larger reddening than M13 by about
$\delta$E(B-V)$\approx$0.02 mag.

We find that adopting the Carretta \& Gratton metallicity value 
for the clusters requires that $\alpha$ cannot be constant when
passing from Z=0.0002 to 0.001 (since for $\alpha$ = 2.0 one obtains 
a good fit for M3 with E(B-V)=0.00, but the fit for M13 requires a negative
reddening). If one increases $\alpha$, the required reddening
increases and Fig 3 (left panel) shows that with the
assumption $\alpha$= 2.3 the TO and subgiant region of M13 are
reasonably well fitted by the isochrone of $\approx$11.5 Gyr with
a distance modulus (m$_{\mathrm v}$ - M$_{\mathrm v}$)$\approx$14.46
and a reddening E(B-V) =0.01$\div$0.02.  This stands 
in good agreement with Carretta et
al. (2000: (m$_{\mathrm v}$ - M$_{\mathrm v}$)= 14.44$\pm$0.04).
However, at larger luminosities, the cluster red giant branch (RGB) 
becomes progressively cooler than predicted.  A similar effect has been
discussed by Brocato et al. (1999) when fitting the RGB of
NGC6362 (whose metallicity is similar to both M3 and M13),
again with $\alpha$=2.3. One may conclude that by adopting such
a large mixing length for MS structures, one cannot avoid 
decreasing mixing length value along the RGB.

This result depends critically on the assumed metallicity.
As shown in the lower panel of Fig.2, decreasing the metallicity
down to the Zinn \& West values makes the isochrone hotter,
increasing the reddening requirements.  In contrast, the SGB
temperature range shrinks, requiring a shorter mixing length. As a result,
the right panel of Fig.3 shows that for Z$\approx$0.0006 M13 can be reasonably
well fitted assuming $\alpha$ = 2.0, with similar values of distance
modulus and reddening, but increasing the age to about 13 Gyr. 
The isochrone now better reproduces the
RGB shape. Thus metallicity plays a critical role in assessing the
calibration of $\alpha$ for intermediate metallicity clusters, such as M13
and M3.  Pending further and more precise constraints on this
parameter, we prefer to adopt $\alpha$ = 2.0 throughout the
explored range of metallicities, both to avoid a peculiar efficiency of
superadiabatic convection at these metallicities and to
better fit the RG branches. However, for the case Z=0.001 we
present in the evolutionary library both cases ($\alpha$ = 2.0 and
$\alpha$=2.3), to allow a quantitative comparison of
uncertainties in the fitting of observed CM diagrams.

Comparison of the two panels in Fig.3 shows 
that the assumed value of $\alpha$ affects the evaluation of the cluster
age, but has negligible impact on the derived distance moduli. We thus 
find that the best fit with the present isochrones appears able to
produce distance moduli of comparable accuracy to those from the
Hipparcos-based evaluations, independent of any uncertainties on the
cluster metallicity.  Rey et
al. (2001) adopt Z=0.0006 to obtain for M13 an age of 13 Gyr, in
agreement with our result. This is the consequence of the
good agreement between present isochrones and those of Yi et al. (2001), 
as we will discuss in the next section and in the Appendix A.

\section{The models}
\begin{figure}[h]
\centering
\includegraphics[width=5.5cm]{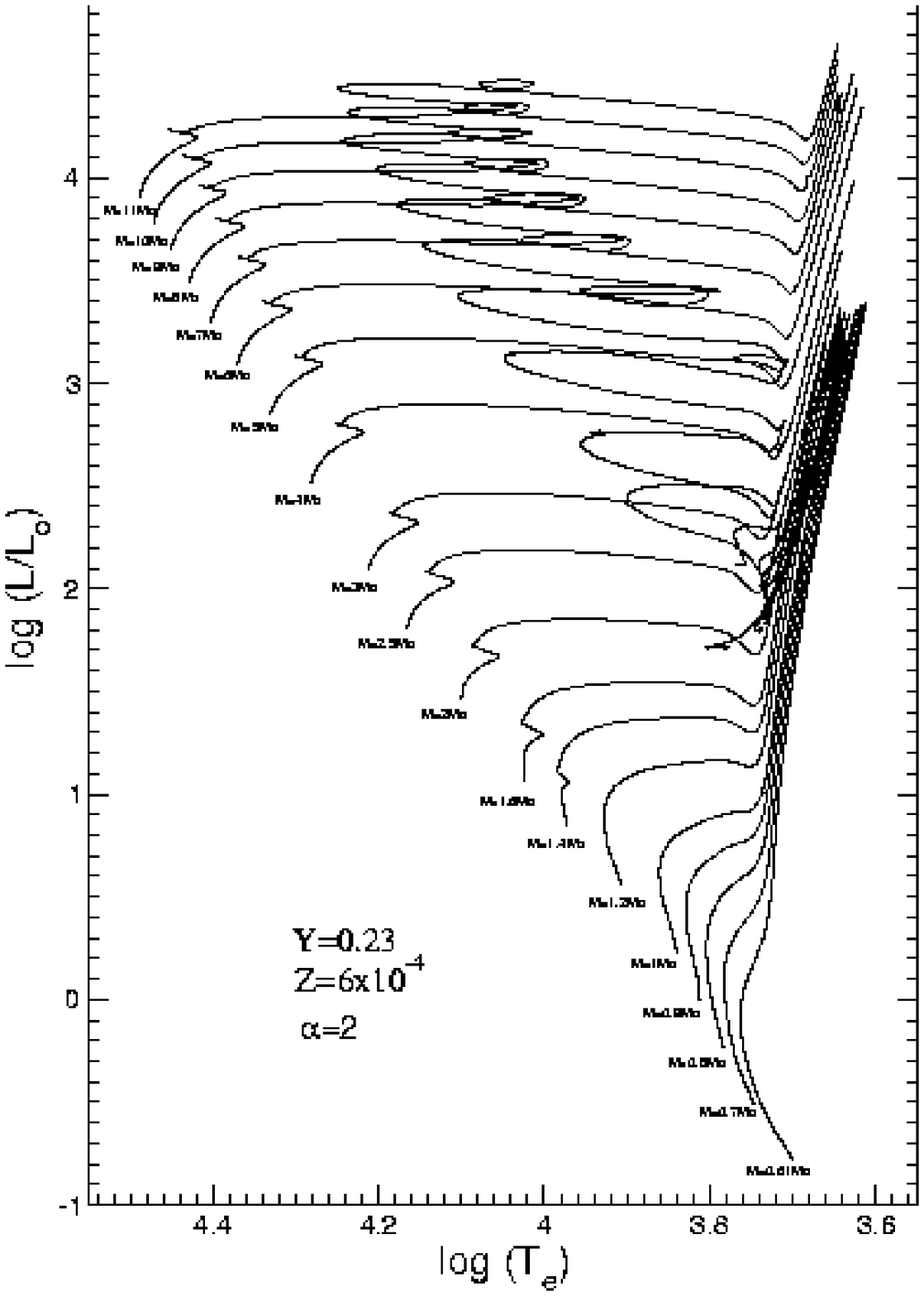}\hspace{0.5cm}\includegraphics[width=5.5cm]{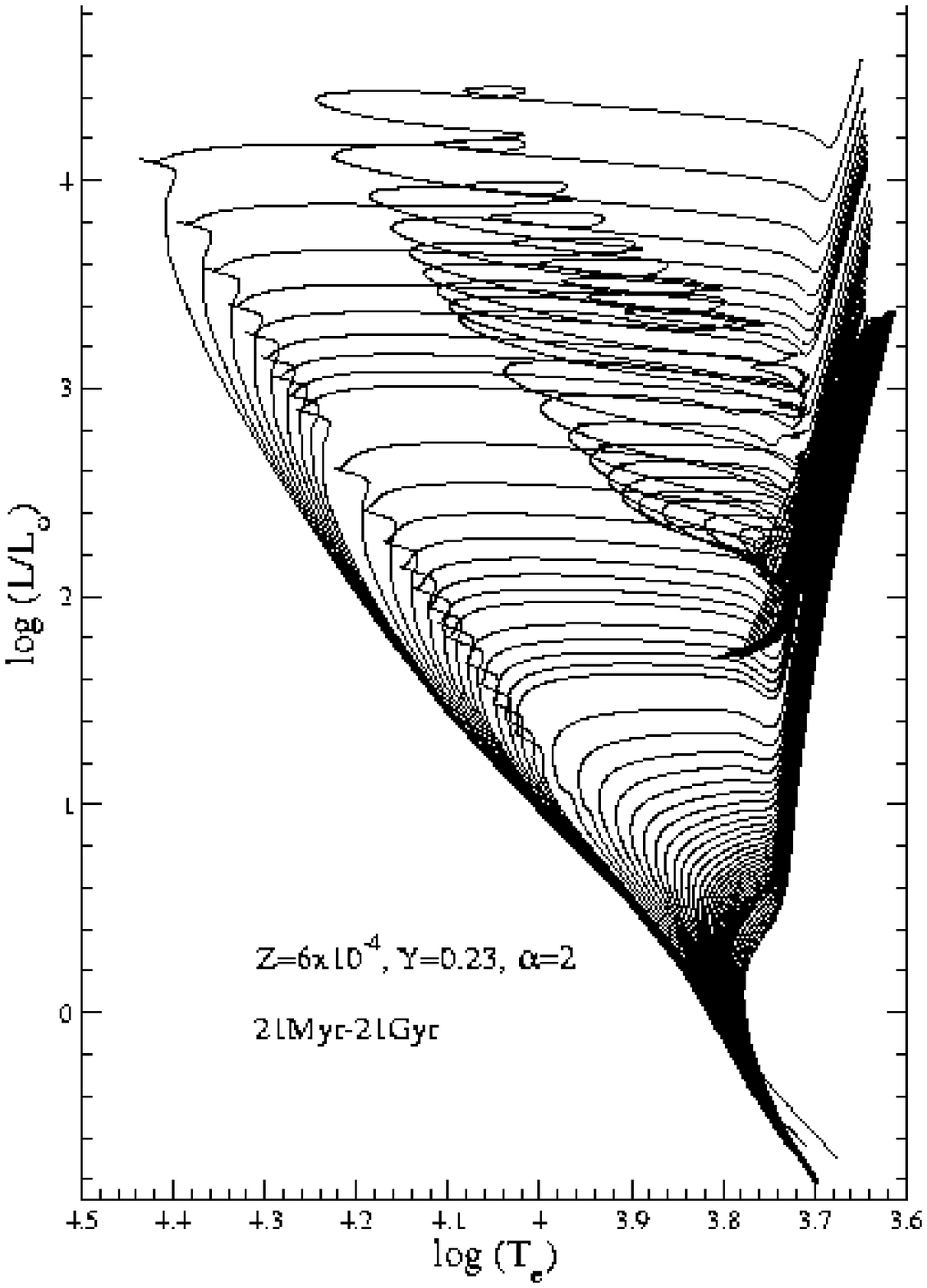}\hspace{0.5cm}\includegraphics[width=5.5cm]{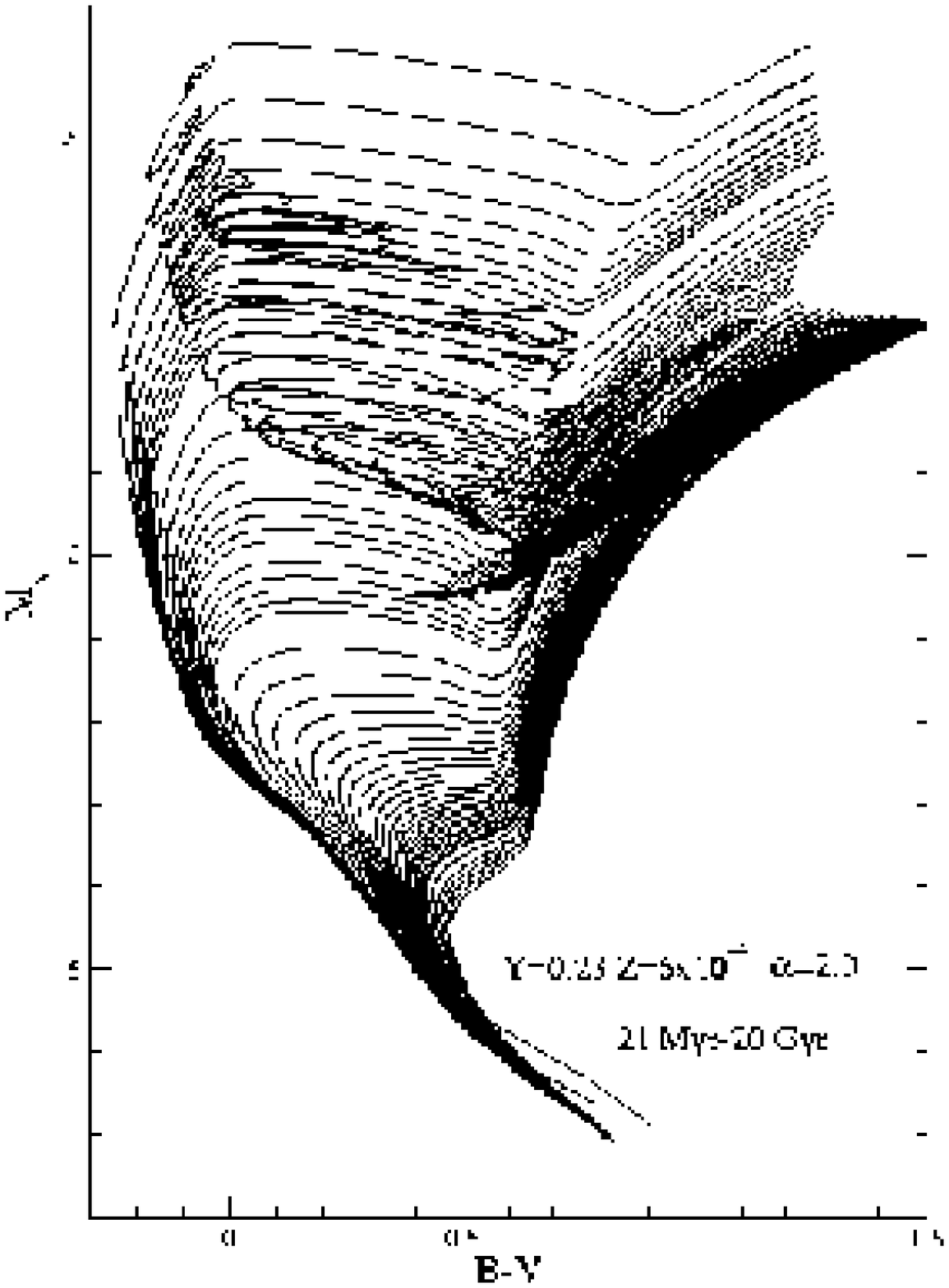}
\caption{Models for Z=0.0006 Y=0.230 $\alpha$=2.0. Left panel:
evolutionary path in the (log(L/L$_{\odot}$),logT$_{\mathrm e}$)
diagram of the models computed for the labelled mass value. Middle
panel: theoretical isochrones in the selected range of ages in the
(log(L/L$_{\odot}$),logT$_{\mathrm e}$) diagram. Right
panel: theoretical isochrones in the selected range of ages in the
(M$_{\mathrm V}$, B-V) diagram.}
\end{figure}
\begin{figure}
\centering
\includegraphics[width=8cm]{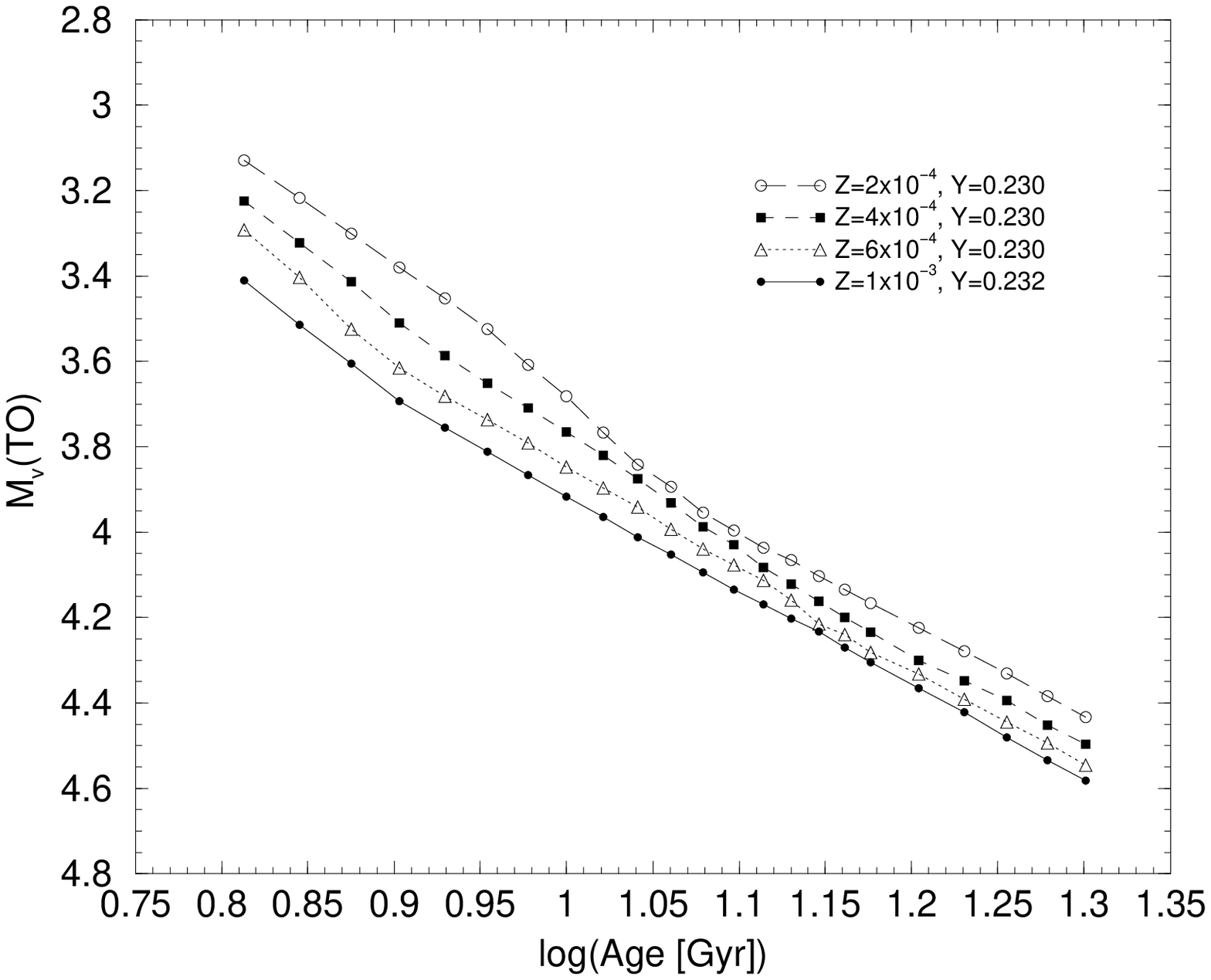}\hspace{0.5cm}\includegraphics[width=8cm]{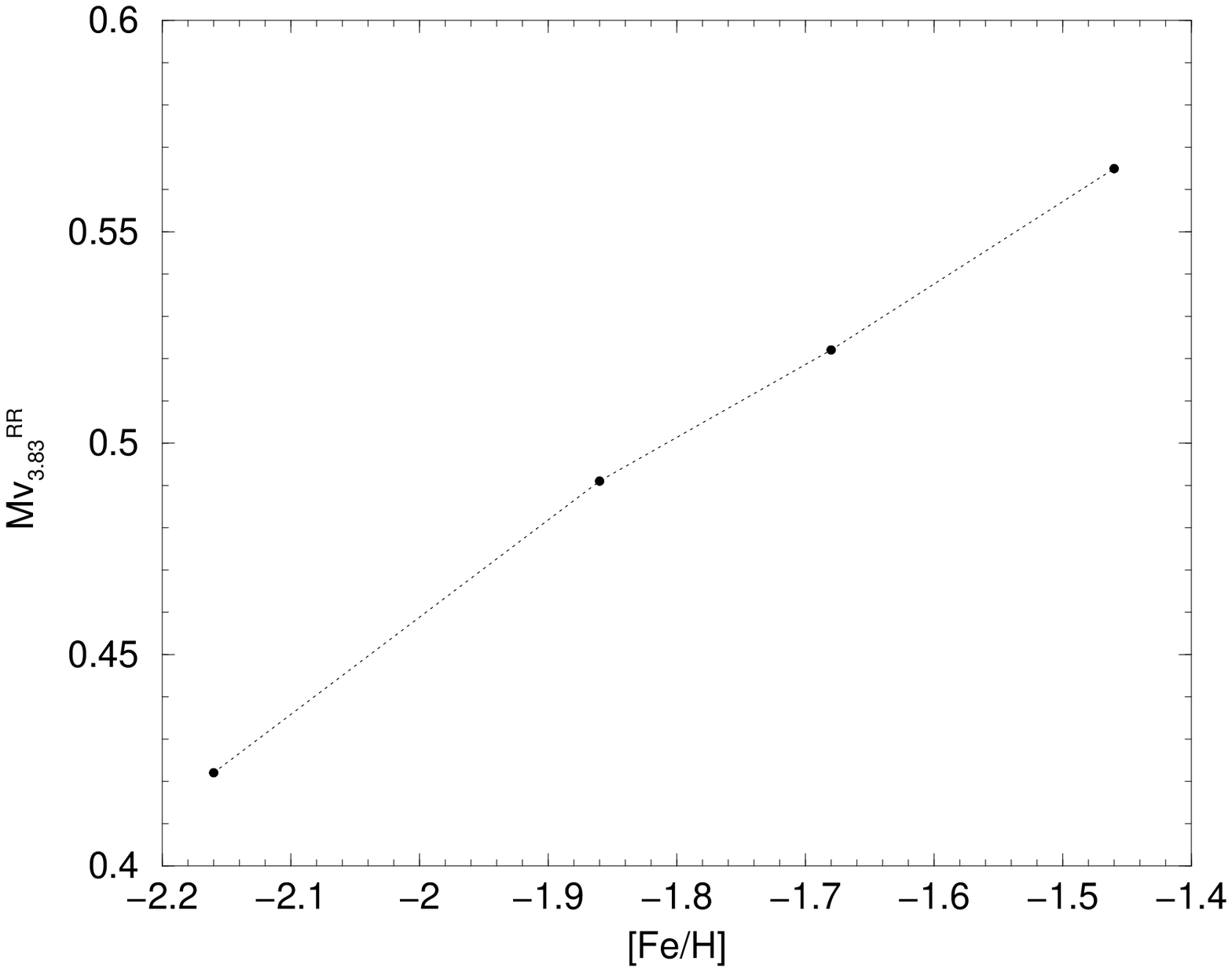}
\caption{Left panel: TO visual magnitude, $M_{\rm V}{\rm (TO)}$, as a
function of the age, for the labeled chemical compositions. Right
panel: ZAHB visual magnitude in the RR Lyrae region (Mv$_{3.83}^{\rm
RR}$) as a function of [Fe/H].}
\end{figure}
\begin{table*}
\caption{The V magnitude of the main sequence termination (MT), the difference in
luminosity between the clump and MT ($\Delta$M$_V$) and the clump mass (m in
M$_{\odot}$) for isochrones with age $\le$ 6 Gyr (see text) and
with the labeled chemical composition.
}
\begin{center}
\begin{tabular}{c c c c c c c c c c c c c c c c}
\hline
\hline
Age & MT & $\Delta$M$_V$  & m &  & MT & $\Delta$M$_V$  & m &  & MT & $\Delta$M$_V$  & m & & MT & $\Delta$M$_V$  & m \\
\hline
\multicolumn{1}{c}{}
&\multicolumn{3}{c}{Z=0.0002 Y=0.23}
&\multicolumn{1}{c}{}
&\multicolumn{3}{c}{Z=0.0004 Y=0.23}
&\multicolumn{1}{c}{}
&\multicolumn{3}{c}{Z=0.0006 Y=0.23}
&\multicolumn{1}{c}{}
&\multicolumn{3}{c}{Z=0.001 Y=0.232}
\\
\hline
 0.15 &   -1.924 &  0.074 & 3.55 & & -1.622 &  0.389 & 3.59  & & -1.628 &  0.359 & 3.63 & & -1.673  & 0.268 & 3.67\\
 0.20 &   -1.232 &  0.328 & 3.11 & & -1.160 &  0.385 & 3.17  & & -1.175 &  0.350 & 3.21 & & -1.228  & 0.258 & 3.25\\
 0.25 &   -0.834 &  0.378 & 2.83 & & -0.809 &  0.385 & 2.88  & & -0.824 &  0.349 & 2.91 & & -0.882  & 0.256 & 2.90\\
 0.30 &   -0.549 &  0.383 & 2.62 & & -0.530 &  0.383 & 2.67  & & -0.547 &  0.347 & 2.70 & & -0.600  & 0.261 & 2.76\\
 0.35 &   -0.382 &  0.332 & 2.47 & & -0.294 &  0.385 & 2.51  & & -0.312 &  0.345 & 2.54 & & -0.354  & 0.271&  2.59\\
 0.40 &   -0.469 &  0.076 & 2.36 & & -0.345 &  0.154 & 2.39  & & -0.267 &  0.208 & 2.42 & & -0.218  & 0.216 & 2.47\\
 0.45 &   -0.550 & -0.154 & 2.27 & & -0.389 & -0.049 & 2.30  & & -0.301 &  0.012 & 2.32 & & -0.223  & 0.043 & 2.36\\
 0.50 &   -0.624 & -0.363 & 2.20 & & -0.431 & -0.235 & 2.22  & & -0.331 & -0.165 & 2.24 & &  -0.229 & -0.113 &2.20\\
 0.60 &   -0.712 & -0.687 & 2.07 & & -0.508 & -0.564 & 2.08  & & -0.383 & -0.473 & 2.10 & & -0.241  & -0.389 &2.12\\
 0.70 &   -0.728 & -0.857 & 1.96 & & -0.538 & -0.759 & 1.97  & & -0.401 & -0.673 & 1.99 & &  -0.270 & -0.613 &2.00\\
 0.80 &   -0.640 & -0.903 & 1.88 & & -0.475 & -0.836 & 1.89  & & -0.346 & -0.769 & 1.90 & & -0.223 &  -0.721 &1.90\\
 0.90 &   -0.559 & -0.942 & 1.81 & & -0.419 & -0.904 & 1.82  & & -0.298 & -0.856 & 1.83 & & -0.183 & -0.820 & 1.84\\
 1.00 &   -0.485 & -0.976 & 1.74 & & -0.368 & -0.965 & 1.76  & & -0.254 & -0.933 & 1.76 & & -0.148 & -0.912 & 1.78\\
 1.50 &   -0.236 & -1.154 & 1.53 & & -0.173 & -1.210 & 1.54  & & -0.099 & -1.226 & 1.55 & & -0.038 & -1.298 & 1.56\\
 2.00 &   -0.121 & -1.354 & 1.40 & & -0.078 & -1.441 & 1.41  & & -0.017 & -1.452 & 1.41 & &  0.017 & -1.528 & 1.42\\
 2.50 &   -0.065 & -1.551 & 1.30 & & -0.018 & -1.633 & 1.31  & &  0.042 & -1.647 & 1.32 & &  0.066 & -1.725 & 1.33\\
 3.00 &   -0.020 & -1.712 & 1.23 & &  0.031 & -1.785 & 1.24  & &  0.090 & -1.802 & 1.24 & &  0.106 & -1.888 & 1.25\\
 3.50 &    0.021 & -1.849 & 1.18 & &  0.072 & -1.922 & 1.18  & &  0.129 & -1.938 & 1.19 & &  0.142 & -2.026 & 1.19\\
 4.00 &    0.058 & -1.956 & 1.13 & &  0.111 & -2.043 & 1.13  & &  0.156 & -2.059 & 1.14 & & 0.163  & -2.151 & 1.15\\
 4.50 &    0.092 & -2.045 & 1.09 & &  0.142 & -2.150 & 1.09  & &  0.180 & -2.161 & 1.10 & &  0.181 & -2.260 & 1.11\\
 5.00 &    0.121 & -2.169 & 1.05 & &  0.170 & -2.239 & 1.06  & &  0.201 & -2.251 & 1.06 & &  0.198 & -2.355 & 1.00\\
 5.50 &    0.148 & -2.249 & 1.02 & &  0.195 & -2.317 & 1.03  & &  0.220 & -2.332 & 1.03 & & 0.213  & -2.442 & 1.05\\
 6.00 &    0.172 & -2.316 & 1.00 & &  0.218 & -2.383 & 1.00  & &  0.238 & -2.411 & 1.01 & &  0.226 & -2.522 & 1.02\\
\hline
\hline
\end{tabular}
\end{center}
\end{table*}
The computed evolutionary models
cover with a fine grid the mass range 0.6 to 11 M$_{\odot}$ 
for the adopted chemical compositions Z=0.0002 Y=0.230, Z=0.0004
Y=0.230, Z=0.0006 Y=0.230, Z=0.001 Y=0.232, where the amount of
original helium has been evaluated by assuming a primordial helium
abundance Y$_P=0.23$ and $\Delta$Y/$\Delta$Z $\sim$2.5 (see e.g. Pagel
\& Portinari 1998, Castellani et al. 1999).  In all cases we assumed
a solar mixture as given by Grevesse \& Noels (1993). 
All the models  with evolutionary times smaller than the Hubble
time have been followed from the Main Sequence through the H
and He burning phases, until the C core ignition or the onset of thermal
pulses in the advanced AGB phase. In the
case of low mass stars undergoing violent He flashes, stellar
structures at the Red Giant Branch (RGB) tip have been used to
produce the corresponding Zero Age Horizontal Branch (ZAHB) models,
further evolving these models until the onset of thermal pulses. Less
massive stars (M$<$0.7~M$_{\odot}$), whose evolutionary times
are longer than the Hubble time, have been evolved up to
central H exhaustion. To allow a proper evaluation of isochrones
with ages lower than $\approx$150 Myr, the Pre-Main Sequence phase
of selected low mass models has been included in the isochrones.

The evolutionary characteristics of the models have
already been described in the literature, and they will not be discussed
here.  We only notice that the mass of the He core at the central helium
ignition appears in good agreement with the results of Dominguez et
al. (1999). On the basis of evolutionary tracks, cluster isochrones in
the age range $\sim$20 Myr to $\sim$20 Gyr have been evaluated for the
four choices about the cluster metallicity; for Z=0.001 models are
calculated for two different $\alpha$ values: 2.0 and 2.3. Figure 5
shows, as an example, the HR diagram for Z=0.0006 $\alpha$=2.0;
isochrones for the two metallicities are also shown in the (M$_{\mathrm
v}$, B-V) plane.

Detailed tables for both tracks and isochrones are available at the
URL http://astro.df.unipi.it/SAA/PEL/Z0.html (and at the CDS).  
For each mass or for each isochronal age, these tables give the luminosity
and effective temperature, followed by the visual magnitude and colors
in the Johnson and infrared Cousins bands, as derived by adopting the
model atmospheres by Castelli (1999). In addition one finds for the
four adopted chemical compositions, tables with the distribution of
ZAHB models as produced by a 0.8 M$_{\odot}$ progenitor giving for the
various HB masses the luminosity, temperature, V magnitude and B-V
color, together with files reporting the HB and AGB evolution of all
these models. This procedure gives ZAHB corresponding to ages in the
range 10 to 14 Gyr, depending on the metallicity. As usual, the small
dependence of these ZAHB luminosities on age can be neglected. A
detailed investigation on the effect of age on ZAHB luminosities has
been published by Cassisi \& Caputo (2002).

To permit a detailed comparison among different evolutionary
computations, the website contains files listing the
 structural evolution of
three selected models (M=0.9, 2.0 and 4.0 M$_{\odot}$) for all the
selected chemical compositions. For each mass, a file
lists the sequence number of the model, its age,
central abundance by mass of H or He, luminosity, effective
temperature, central temperature and density, the maximum
off-center temperature and its location, the mass of the
convective core, He core and convective envelope and, in the last
four columns, the fraction of the total luminosity released by pp,
CNO and He nuclear burning and by the gravothermal energy.

The evolutionary sequences  for old clusters with ages greater than
6 Gyr closely follow theoretical predictions discussed in
Cassisi et al. (1998, 1999). Here we take advantage of the finer  grid
of metallicities to present in Fig.5, left panel,  an
improved calibration of the Turn Off magnitudes as a function of
the cluster age covering the whole range 0.0002$\leq$ Z $\leq$ 0.001.
The right panel in Fig.5 gives the calibration of the ZAHB
magnitude taken at logT$_{\mathrm e}$=3.83, i.e., near the central region of
the RR Lyrae instability strip as a function of the cluster
metallicity. A linear best fit through these data, taking into account
an $\alpha$ enhancement of [$\alpha$/Fe]$\approx$0.3, gives:
\begin{displaymath} 
M_V (ZAHB) = 0.20 [Fe/H] + 0.86
\end{displaymath}
a result well within current observational constraints (e.g.
Chaboyer et al. 1998):
\begin{displaymath}
M_V (ZAHB) = (0.23 \pm 0.04) [Fe/H] + (0.83 \pm 0.08)
\end{displaymath}
or by Carretta et al. (2000):
\begin{displaymath}
M_V (ZAHB) = (0.18 \pm 0.09) [Fe/H] + (0.74 \pm 0.12)
\end{displaymath}
Unfortunately, the large uncertainties in the empirical relations do not permit
a solution to the problem of a possible over-luminosity of present theoretical predictions.

As well known (see e.g. Stetson et al. 1996, Salaris \& Weiss 1997,
Cassisi et al. 1998, 1999) TO and ZAHB magnitudes can be combined to
calibrate the difference in magnitude ($\Delta$M$_V$) between the
Turn-Off (TO) and the helium burning HB phase as an age indicator,
independently of both cluster distance and reddening. As discussed in
Paper I, for ages $\leq$6 Gyr the difference between the bright
end of the main sequence (MT) and the bottom 
luminosity of the clump of He burning
giants can provide an observational parameter that is able to derive
from the CM diagram at least a rough estimate of the age
independently on the cluster distance (see
e.g. Castellani et al. 1999).  Table 2 gives the V magnitude of the main sequence termination (MT) and
the difference ($\Delta$M$_V$) in visual magnitude between the clump and
MT as a function of the age in the range 150 Myr to 6 Gyr for all the
metallicities investigated in this paper. MT and the clump magnitude
have been evaluated at the maximum luminosity reached just after the
overall contraction (H exhaustion) and at the minimum luminosity of
the He clump region, respectively. For each age, the original mass of
stars populating the He burning clump is also shown. In the case of old
globulars, our calibration gives ages in reasonable agreement with the ages
recently determined by  Salaris \& Weiss (2002). However, with
respect to Salaris \& Weiss the present isochrone database has
the additional advantage of including  ages much lower than the
typical globular clusters ages,  down to about 20 Myr.
\begin{figure}
\centering
\includegraphics[width=8cm]{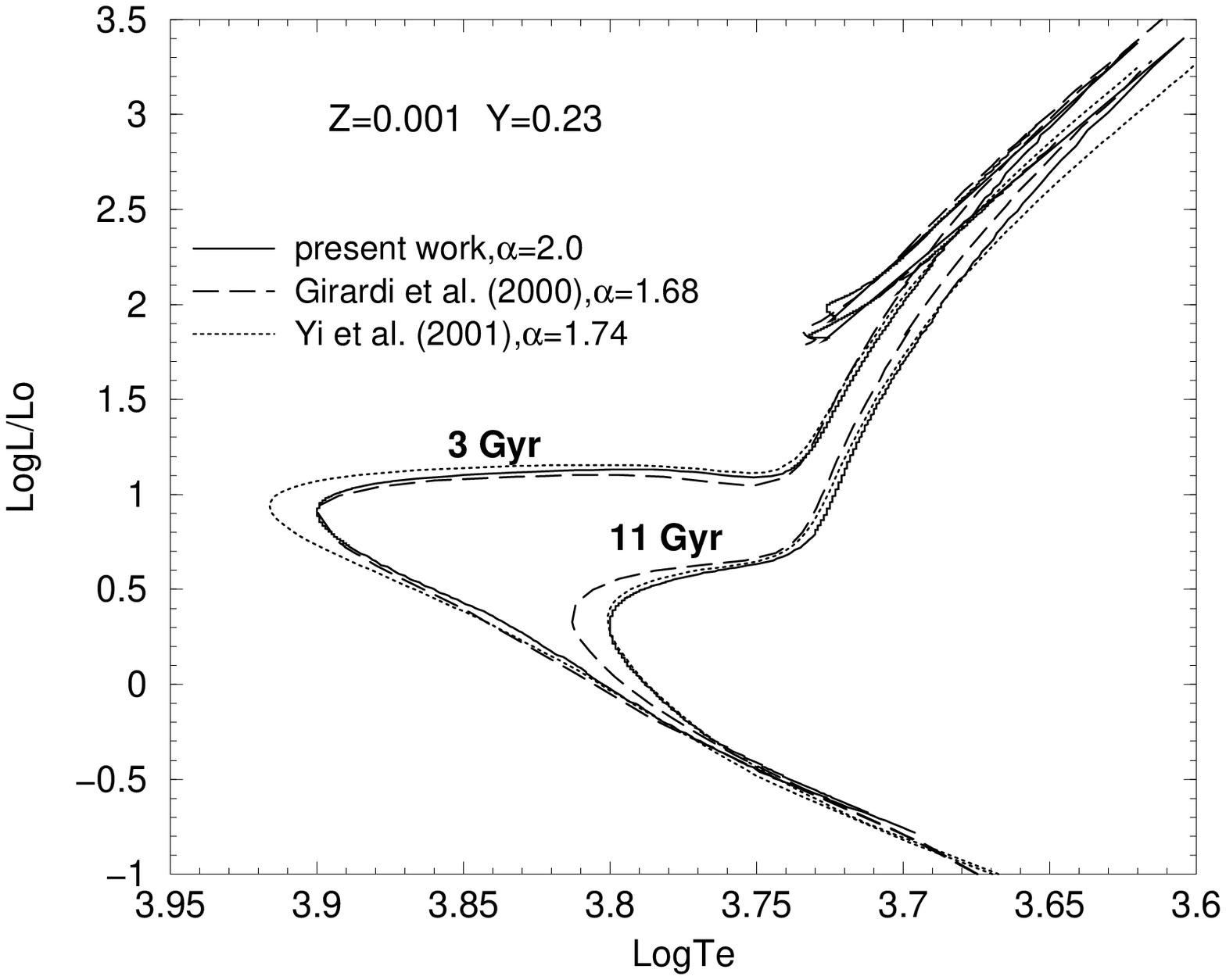}\includegraphics[width=8cm]{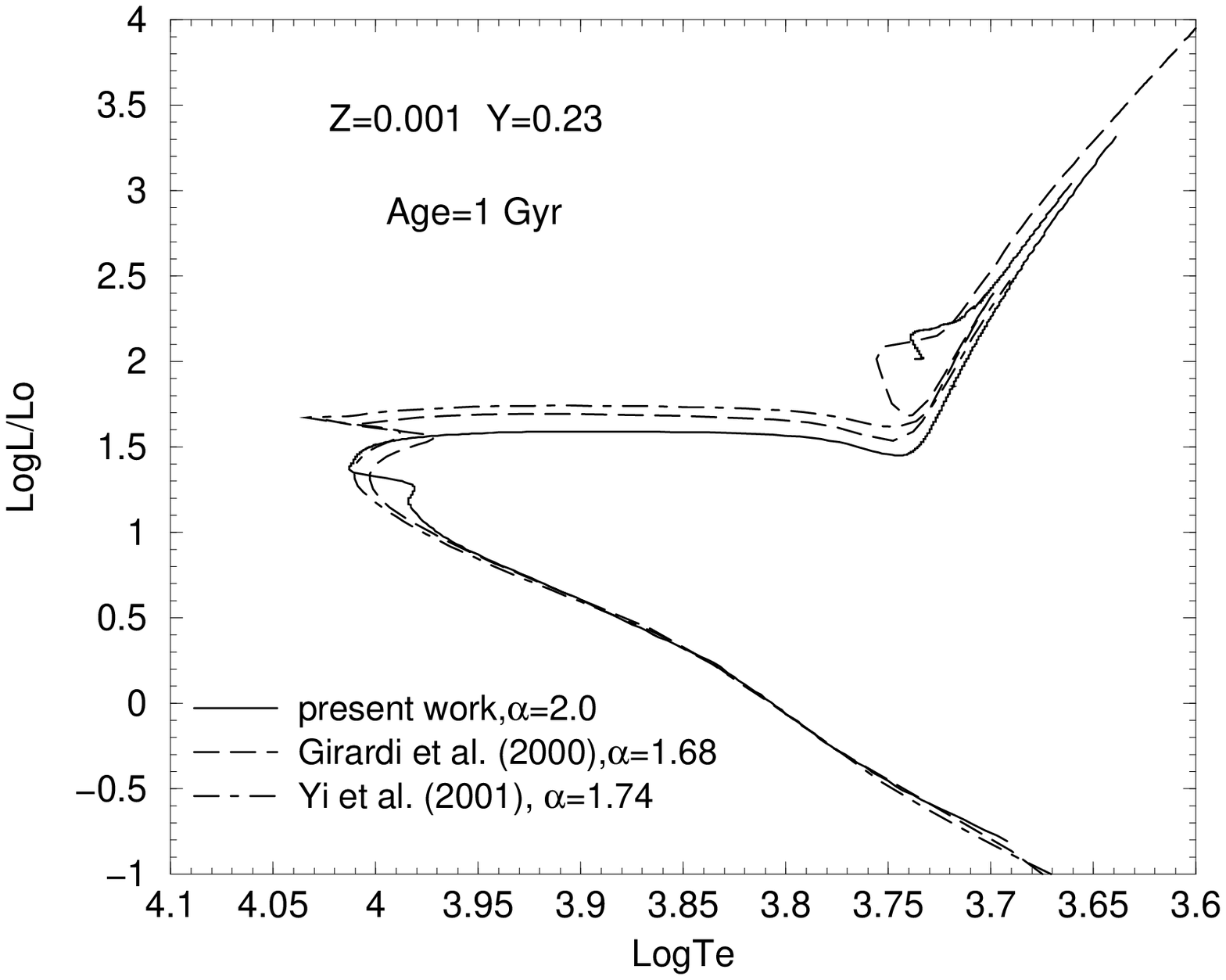}
\caption{Left panel: comparison between present, Girardi et al. (2000)
and Yi et al.(2001) isochrones in the HR diagram for Z=0.001 Y=0.23
and for the two different labeled ages. Righ panel: as in the left panel
but for age=1 Gyr.}
\end{figure}

\section{Comparison with previous results}

A discussion of the differences between the present and
Girardi et al. (2000)
 evolutionary models has already been given in Paper I
(see also Castellani et al. 2000), whereas the existing differences
between Girardi et al. and Yi et al. (2001) have been discussed in
Gallart et al. (2003).  Fig.6 shows a comparison between selected
isochrones  in the theoretical (logL/L$_{\odot}$, logT$_{\mathrm e}$) plane for
Z=0.001. A broader discussion on the main characteristics of the most
updated extended sets of evolutionary models available in the literature
can be found in Appendix A. 

Inspection of data in Fig.6 (left panel) shows that in the range of
the largest ages (age$\approx$10 Gyr, mass at the TO
M$\approx$0.85M$_{\odot}$) isochrones without element diffusion,
e.g. in Girardi et al. (see also Salasnich et al. 2000), for any given
TO luminosity give larger cluster ages and larger TO temperatures than
diffusion isochrones.  One finds a good agreement with the models
of Yi et al. along the MS and SGB as a result of the quite similar
input physics adopted to compute H burning models; the differences in
temperatures in the upper portion of the RGB branch likely originate
from different assumptions about the mixing length parameter $\alpha$.

Differences in temperatures between our models and Yi et al., both
allowing for element diffusion, originate from
different assumptions about the mixing length parameter $\alpha$, and
one finds that our large age isochrones with $\alpha$= 2.0 agree well
with the ones of Yi et al. along the MS and SGB as a result of the
quite similar input physics adopted to compute H burning models.

For lower ages, one notes that the TO.s of the 3 Gyr isochrones with
(Girardi et al. 2000, Yi et al. 2001) or without (present
computations) overshooting appear rather similar, with negligible
differences between our and the Girardi et al. results.  The
reasons for this are twofold: the overshooting values adopted by Yi et
al. and Girardi et al. are lower than those originally adopted in
the Padua isochrones (see e.g. Bertelli et al. 1986), moreover for
1.0$\lcu$M/M$_{\odot}\lcu$1.5 the overshooting efficiency
is further reduced, and for M/M$_{\odot}\leq$1.0 the overshooting
amount is set to zero to avoid the development of a small central
convective zone in the solar model which would persist up to the
present age of 4.6 Gyr.  Note that the difference in TO temperature
between models with and without overshooting does not scale as
expected with the assumptions on $\alpha$, due perhaps to different
treatments of the superadiabatic convection/inclusion of
undershooting. As already discussed in Castellani et al. (2000)
Girardi et al. models give fainter He burning clumps.  We conclude
that, due to the reduced extension of the convective
cores of the stars populating clusters with Z$\approx$0.001 and ages
larger than or of the order of 3 Gyr, these clusters cannot provide
any evidence for or against overshooting.

Evidence for efficient overshooting can however be found in younger
clusters, with ages around 1 Gyr, as shown in the right panel of
Fig.6; one finds that at this age overshooting pushes the model beyond
the Red Giant transition, and the overshooting signature will be
represented by the reduced extension and population of the RG
branch.  We identify this prediction as the most robust test for
overshooting, as already discussed in the seminal paper by Barbaro \&
Pigatto (1984). However such a test is
sensitively dependent on the cluster metallicity. A similar comparison
of available isochrones reveals, for instance, that clusters 
at Z=0.004 with ages greater than about 1.5 Gyr, as the ones discussed
in the recent paper by Woo et al. (2003), do not provide information 
about the overshooting efficiency.

\section{Final remarks}

In this paper we present and discuss canonical stellar evolutionary
models produced for a grid of chemical compositions covering the range
of Population II stars. For each given metallicity, we adopted a
mixing length calibrated to reproduce the observed color of red giant
branch in galactic globulars. On this basis we make available
evolutionary tracks and cluster isochrones covering the range of ages
20 Myr to 20 Gyr.  The difference in magnitude between the top MS (the
Blue Sequence) and He burning structures has been finally calibrated
in terms of the cluster age for globular and open clusters.  Due to
the wide range in age, this calibration can be applied not only to
population II clusters in the Galaxy but also to intermediate and
young, low metallicity clusters in dwarf galaxies and in the
Magellanic Clouds. VandenBerg, Bolte \& Stetson (1990) cautioned that 
uncertainties in input physics and/or model atmospheres make it unlikely
that the predicted color difference between RG and TO is
quantitatively correct.  The discussion presented in this paper shows
that a correct description of this parameter appears now
within the theoretical possibilities as a result of the recent
improvement in both input physics and atmospheric models.
  
\begin{acknowledgements}
We warmly thank Steve Shore and Pier Giorgio Prada Moroni
for a careful reading of the manuscript and the anonymous referee
for very useful comments.
Financial support for this work was provided by the Ministero
dell'Istruzione, dell'Universit\`a e della Ricerca (MIUR) under
the scientific project ``Stellar observables of cosmological
relevance'' (V. Castellani \& A. Tornamb\`e, coordinators).
\end{acknowledgements}

\appendix {
\section{Stellar evolutionary models}
Due to the relatively large number of stellar models available in the 
literature, we discuss the current situation.
Theoretical predictions concerning
stellar structures and their evolution depend on the adopted
microphysics describing the behavior of the stellar plasma as well as
on the assumptions about the efficiency of physical processes such as
convection, microscopic diffusion, and overshooting. The comparison with
observations, which is the final goal of evolutionary computations,
requires a third ingredient: the model atmospheres needed
to translate theoretical luminosities and effective temperatures into
magnitudes and colors.
\begin{table*}
\scriptsize
\caption{Comparison of the main characteristics of the most recent
extended sets of evolutionary models available in the literature, see
text.}
\begin{center}
\begin{tabular}{lccccc}
\hline
\hline\\
Models      & Present & Yi et al. 2001 & Girardi et al. 2000 & Geneva & VandenBerg et al.2000\\
\hline \hline

Mass Range [M$_{\odot}$] &0.6-11 & 0.4-5 & 0.15-7 & 0.8-120 & 0.5-1.0 \\
Metallicity (Z) &$2\cdot10^{-4}\div8\cdot10^{-3}$ &  $10^{-5}\div8\cdot10^{-2}$ & $4\cdot10^{-4}\div3\cdot10^{-2}$* & $10^{-3}\div10^{-1}$& $10^{-4}\div3\cdot10^{-2}$*\\
Evo.Phases  & H+He & H & H+He & H+He & H+He\\
\hline
$\alpha$       & 1.9-2.3 & 1.74 & 1.68 & 1.6 & 1.89 \\
Diffusion   & Thoul et al. 1994 &Thoul et al. 1994* & NO & NO & NO*\\
Overshooting  & NO* & 0.2 Hp & 0.25 Hp & 0.20Hp & - \\
\hline
EOS         & OPAL96+Str.1998*  & OPAL96+Yale      & Padua+MHD       & Geneva*      & Victoria \\
Rad. Opacity     & OPAL96+Alex.1994* & OPAL96+Alex.1994 &OPAL92+Alex.1994 &OPAL92+Kur.1991 & OPAL96+Alex.1994 \\
Nucl.Reactions & NACRE* & Bahcall\&P.1992 & Caug.\&Fow.1988 & Caug.\&Fow.1988 & Bahcall\&P.1992\\
El. Conduction  & Itoh et al.1983 &  Hubb.Lampe1969* & Hubb.Lampe 1969 & Hubb.Lampe1969 & Hubb.Lampe1969 \\
Neutrinos   & Itoh et al.1996 & Itoh et al.1989 & Munakata et al.1985 & Itoh et al.1989 & Itoh et al.1996\\
\hline
Model Atm.  & Castelli et al.1999 & Lejeune et al.1999* & Castelli et al.1999* & Lejeune et al.1999 & Vand.\& Clem2003* \\
\hline
\hline
\end{tabular}
\end{center}
\end{table*}

Current stellar models differ in
a variety of choices concerning these inputs, producing small but
significant differences in the results.  Table 
A1 summarizes the main features of the 
sets of models (tracks and isochrones) available in the
literature, which, to our knowledge, cover a wide range in
chemical composition and stellar masses/isochrones ages.
The quoted models obviously do not exhaust the rich and composite
scenario of updated stellar evolutionary calculations; for example
several authors calculated sets of tracks/isochrones for ancient
populations or for very low mass stars (see e.g. Straniero et
al. 1997, Chabrier \& Baraffe 1997, Cassisi et al. 2000, Vandenberg et
al. 2000) or analyzed the characteristics of sets of intermediate mass
stars (see e.g. Dominguez et al. 1999, Bono et al. 2000). In the table
we add only  models by VandenBerg et al. (2000) for low mass
stars, as an example of evolutionary grids for old
ages. 

The table shows in the first three lines the range of masses, the range
of metallicities (Z) and the evolutionary phases covered by the quoted
models. The following three lines give information on the treatment of
the macroscopic mechanism, listing the adopted values of the mixing
length parameter governing the efficiency of superadiabatic convection
($\alpha = l/H_P$) and indicate whether or not diffusion and/or
overshooting have been taken into account. The following five lines
give the sources of the adopted physical inputs,
listing in order the sources for the Equation of State (EOS),
radiative opacity, cross sections for nuclear reactions, 
the evaluation of the electron conduction in degenerate matter and
the production of cooling neutrinos. The adopted atmospheric models
are listed in the last row.  Asterisks in the table mark the data
that require some additional information.

Present models: computations presented in this paper for the
metallicity range Z= 0.0002 to 0.001 do not include overshooting.
 However, the optional occurrence of overshooting has
been presented and discussed in Paper I for the cases Z=0.004 and
0.008. OPAL96 EOS (Rogers et al. 1996) and OPAL96 Radiative Opacity
(Iglesias \& Rogers 1996) when needed have been implemented with
an updated version of the EOS of Straniero (1988) and with low
temperature opacities from Alexander \& Fergusson (1994),
respectively. Reaction rates are from the NACRE collaboration (Angulo
et al. 1999). \\

Yi et al.: computations with different choices about the overabundance
of ``$\alpha$" elements ([$\alpha$/Fe]= 0.0, 0.3, 0.6) are
presented. For the conductive opacity Hubbard \& Lampe (1969) is
adopted for log $\rho$(in g/cm$^3$)$\leq 6.0$ and Canuto (1970) for
log$\rho$(in g/cm$^3$)$>6.0$.  Diffusion is taken into account only
for He. Magnitudes and colors are available as calculated by adopting
two different atmospheric models: either from an update of Green et
al. (1987) or from Lejeune et al. (1998) and Westera et al.  (1999).\\

Girardi et al.: $\alpha$ enhanced models are presented elsewhere
(Salasnich et al. 2000) for some metallicity values. The
original Padua EOS has been implemented with MHD (Mihalas et al.
1990 and references therein) for temperatures lower than 10$^7$ K.
For a critical analysis of the adopted
color transformations in several photometric systems see Girardi
et al. (2002).\\

Geneva: we refer to a series of papers from Schaller al.  (1992) to
Mowlavi et al.(1998).\\

VandenBerg et al.: in more recent models (VandenBerg et al. 2002) both
gravitational settling and radiative acceleration are 
taken into account. We quoted VandenBerg et al. (2000) 
because in VandenBerg et al. (2002) HB models are not calculated.  The 
original helium abundance is
assumed following the formula: Y=0.235 + 2Z. Computations for
models with an overabundance of``$\alpha$" elements ([$\alpha$/Fe]=0.3, 0.6) 
are presented too. The color transformations adopted in
the paper are quoted as ``VandenBerg 2000, in preparation''; 
VandenBerg \& Clem (2003) is probably the correct reference for 
the published paper.\\}

\end{document}